\documentclass[a4paper,hidelinks]{article}
\usepackage{url,bm,algorithm,algpseudocode,mathtools,bbm,pgfplotstable,bm,amsmath,amssymb,amsthm,amsfonts,graphics,graphicx,epsfig,rotating,caption,subcaption,natbib,appendix,titlesec,multicol,hyperref,verbatim,bbm,algorithm,algpseudocode,pgfplotstable,threeparttable, booktabs,mathtools,dsfont}

\newcommand{\x}{\mathbf{x}}
\newcommand{\bbeta}{\bm{\beta}}
\newcommand{\X}{\mathbf{X}}
\newcommand{\y}{\mathbf{y}}
\newcommand{\G}{\mathcal{G}}
\newcommand{\btau}{\bm{\tau}}
\newcommand{\blambda}{\bm{\lambda}}
\newcommand{\bomega}{\bm{\omega}}
\newcommand{\btheta}{\bm{\theta}}
\newcommand{\bmu}{\bm{\mu}}
\newcommand{\bSigma}{\bm{\Sigma}}
\newcommand{\bpsi}{\bm{\psi}}
\newcommand{\bOmega}{\bm{\Omega}}
\newcommand{\Z}{\mathbf{Z}}
\newcommand{\bLambda}{\bm{\Lambda}}
\newcommand{\m}{\mathbf{m}}

\newcommand{\tr}{^{\text{T}}}
\newcommand{\expit}{\text{expit}}
\newcommand{\argmax}{\text{argmax} \,}
\renewcommand{\L}{\mathcal{L}}
\newcommand{\E}{\mathbb{E}}
\newcommand{\diag}{\text{diag}}

\renewcommand{\(}{\left(}
\renewcommand{\)}{\right)}
\renewcommand{\[}{\left[}
\renewcommand{\]}{\right]}
\newcommand{\norm}[1]{\left\lVert #1 \right\rVert}

\title{Adaptive group-regularized logistic elastic net regression}
\date{}
\author{Magnus M. M\"unch$^{1,2}$\footnote{Correspondence to: \href{mailto:m.munch@vumc.nl}{m.munch@vumc.nl}}, Carel F.W. Peeters$^{1}$, Aad W. van der Vaart$^{2}$, and \\ Mark A. van de Wiel$^{1,3}$}

\begin{document}
	
	\maketitle
	
	\noindent
	1. Department of Epidemiology \& Biostatistics, Amsterdam Public Health research institute, VU University Medical Center, PO Box 7057, 1007 MB
	Amsterdam, The Netherlands\\
	2. Mathematical Institute, Leiden University, Leiden, The Netherlands \\
	3. Department of Mathematics, VU University, Amsterdam, The Netherlands \\
	
	\begin{abstract}
		{In high-dimensional data settings, additional information on the features is often available. Examples of such external information in omics research are: (a) $p$-values from a previous study, (b) a summary of prior information, and (c) omics annotation. The inclusion of this information in the analysis may enhance classification performance and feature selection, but is not straightforward in the standard regression setting. As a solution to this problem, we propose a group-regularized (logistic) elastic net regression method, where each penalty parameter corresponds to a group of features based on the external information. The method, termed \texttt{gren}, makes use of the Bayesian formulation of logistic elastic net regression to estimate both the model and penalty parameters in an approximate empirical-variational Bayes framework. Simulations and an application to a colon cancer microRNA study show that, if the partitioning of the features is informative, classification performance and feature selection are indeed enhanced.}
	\end{abstract}

	
	\noindent\textbf{Software available from}: \url{https://github.com/magnusmunch/gren/}\\
	
	\section{Introduction}
	Prediction from high-dimensional data is a challenge common to many fields of research. Examples of such high-dimensional prediction problems arise in computer vision, stock market prediction and disease diagnosis from omics data. In this paper, we specifically focus on the latter setting. High dimensionality of data introduces several issues in the estimation of prediction models, especially: (a) unidentifiable models, (b) highly correlated predictor variables and (c) non-trivial selection of variables.
	
	Attempts to tackle one or more of these issues have lead to the development of many prediction methods \cite[]{fan_selective_2010}. Here we will focus on the penalization approach. A well-known and popular penalization method is elastic net regression \cite[]{zou_regularization_2005}, with its special cases ridge regression \cite[]{hoerl_ridge_1970} and lasso regression \cite[]{tibshirani_regression_1996}. The elastic net approach takes issues (a)-(c) into account and has yielded many successful extensions. Recent work on elastic net regression (and its special cases) has focussed on increasing prediction accuracy by the inclusion of prior knowledge on the variables \cite[]{van_de_wiel_better_2016,tai_incorporating_2007}.
	
	In omics research, sources of such prior knowledge on the variables are often available. The information can, for example, come in the form of (a) results on the same molecular features obtained in a previous study (e.g. $p$-values), (b) information from a publicly available database that summarizes the prior information on the molecular features involved (e.g., the Cancer Gene Census \cite[]{futreal_census_2004}), (c) omics annotation (e.g. the location of a gene on the chromosome) and (d) response-independent summary statistics, such as sample standard deviation of the features. Although this information can rarely be directly included in the statistical analysis, it may still be useful and informative for the study at hand. 
	
	A natural way of including these information types is through differential penalization: that is, each group of variables receives its own penalty parameter. An apparent issue with this differential penalization is the estimation of the penalty parameters. Naive estimation may be done by cross-validation. However, cross-validation requires re-estimation of the model over a grid of penalty parameters. The size of this grid increases exponentially with the number of penalty parameters. Consequently, it quickly becomes computationally infeasible. We therefore propose an efficient alternative: empirical Bayes estimation of the penalty parameters, which corresponds to hyperparameter estimation in the Bayesian prior framework. Because of the ubiquity of binary outcome data in omics research, we focus on the logistic elastic net.
	
	We introduce the Bayesian logistic (generalised) elastic net model in Section \ref{sec:model}. In Section \ref{seq:estimation} we derive a variational approximation to this model and use this novel approximation in the empirical Bayes estimation of multiple, group-specific penalty parameters. In Section \ref{sec:extensions} we introduce some extensions of the method, including, among others, an extension to unpenalized covariates. In Sections \ref{sec:simulations}, and \ref{sec:app}, we demonstrate the approach in a simulation study and a microRNA colon cancer data example, respectively. In Section \ref{sec:discussion} we conclude by both discussing some differences and parallels between the proposed method, termed \texttt{gren}, and other methods in the literature.
	
	\section{Model}\label{sec:model}
	\subsection{Logistic elastic net regression}\label{sec:logitmodel}
	In logistic regression the outcome variables are assumed to be binary or sums of $m_i$ disjoint binary Bernoulli trials ($y_i = \sum_{l=1}^{m_i} k_l, k_l \in \{0,1\}$ for $i=1, \dots, n$). The binomial logistic model relates the responses to the $p$-dimensional covariate vectors $\x_i = \begin{bmatrix} x_{i1} & \cdots & x_{ip}\end{bmatrix} \tr$ through $y_i \sim \mathcal{B} \( m_i, \expit ( \x_i\tr \bbeta ) \)$, where $\mathcal{B} (m,\upsilon)$ is the binomial distribution with number of trials $m$ and probability $\upsilon$, and $\expit \( \x_i\tr \bbeta \) = \exp(\x\tr_i \bbeta)/[1 + \exp(\x\tr_i \bbeta)]$. Note that if $m_i=1$ for $i=1, \dots, n$, the model reduces to a binary logistic regression model. Throughout the rest of the paper we assume that the model matrix $\X = \begin{bmatrix} \x_1 & \cdots & \x_n \end{bmatrix}\tr$ is standardized such that $\frac{1}{n}\sum_{i=1}^n x_{ij}=0$ and $\frac{1}{n}\sum_{i=1}^n x_{ij}^2=1$ for $j=1, \dots, p$.
	
	In elastic net regression, the penalised likelihood is maximised to yield parameter estimates:
	$$
	\hat{\bbeta} := \underset{\bbeta}{\argmax} \ell(\y ; \bbeta) - \frac{\lambda_1}{2} \norm{\bbeta}_1 - \frac{\lambda_2}{2} \norm{\bbeta}_2^2,
	$$
	with $\lambda_1, \lambda_2 \in \mathbb{R}_{>0}$ and $\y = \begin{bmatrix} y_1 & \cdots & y_n \end{bmatrix} \tr$. The likelihood term is thus complemented with a penalty term consisting of the $L_1$ and $L_2$-norm of the parameters, multiplied by respective penalty parameters $\lambda_1$ and $\lambda_2$. We have scaled the penalty parameters by $\frac{1}{2}$, to make them correspond with the Bayesian version in the following. The maximiser of the penalised likelihood corresponds to the posterior mode of a Bayesian elastic model with prior \cite[]{li_bayesian_2010,zou_regularization_2005}:
	\begin{equation}\label{eq:elasticnetprior}
	\beta_j \sim g(\lambda_1,\lambda_2) \cdot \exp\[-\frac{1}{2} (\lambda_1 |\beta_j| + \lambda_2 \beta_j^2)\].
	\end{equation}
	Here, $g(\lambda_1,\lambda_2)$ denotes a normalizing constant that is dependent on $\lambda_1$ and $\lambda_2$ (given in Section 2 of the Supplementary Material (SM), along with further details on the elastic net prior). The elastic net combines $L_1$ and $L_2$-norm penalisation such that the model parameters are shrunken towards zero. The $L_1$-norm may set some of the estimates exactly to zero, thus automatically selecting features. The $L_2$-norm ensures that correlated features behave similarly, as one would generally require. We extend the elastic net model to the generalised elastic net to allow for different penalty parameters per group of features. 
	
	\subsection{The generalised elastic net}\label{sec:genen}
	Assume we have a partitioning of the features in $G$ groups, such that each feature belongs to one group. Let $\G(g)$ be the feature index set of group $g$ for $g=1, \dots, G$. To apply differential penalisation we include a group-specific weight $w_g \in \mathbb{R}_{>0}$. The model parameters are now estimated by:
	\begin{align}
	\hat{\bbeta} & := \underset{\bbeta}{\argmax} \ell(\y ; \bbeta) - \frac{\lambda_1}{2} \sum_{g=1}^G \sum_{j \in \G(g)} |w_g \cdot \beta_j| - \frac{\lambda_2}{2} \sum_{g=1}^G \sum_{j \in \G(g)}\( w_g \cdot \beta_j\)^2 \label{eq:weightedlik}\\ 
	& = \underset{\bbeta}{\argmax} \ell(\y ; \bbeta) -  \frac{\lambda_1}{2} \sum_{g=1}^G \sqrt{\lambda'_g} \sum_{j \in \G(g)} | \beta_j| - \frac{\lambda_2}{2} \sum_{g=1}^G \lambda'_g \sum_{j \in \G(g)} \beta_j^2, \nonumber
	\end{align}
	where we wrote $w^2_g = \lambda'_g$ to emphasise that these group-specific weights may be interpreted as penalty multipliers. Throughout the following we assume that the geometric mean of the multipliers, weighted by their respective group sizes, is one, such that the average shrinkage of the model parameters is determined by the `global' $\lambda_1$ and $\lambda_2$. That is, we calibrate the $\lambda'_g$ such that $\prod_{g=1}^G (\lambda'_g)^{|\G(g)|}=1$. The multiplier appears in square root form in the $L_1$-norm term to ensure that penalisation on the parameter level is the same for the $L_1$ and $L_2$-norm terms, as can be seen from the parametrisation in (\ref{eq:weightedlik}). 
	
	\cite{li_bayesian_2010} show that (\ref{eq:elasticnetprior}) may be written as a computationally more convenient scale mixture of normals, with mixing parameter $\btau = \begin{bmatrix} \tau_1 & \cdots & \tau_p \end{bmatrix} \tr$. Using this result, we write the generalised elastic net model in its Bayesian form as:
	\begin{subequations}\label{eq:enmodel2}
		\begin{align}
		\y| \bbeta & \sim \prod_{i=1}^n \mathcal{B} \left( m_i, \expit ( \x\tr_i \bbeta ) \right),  \\
		\bbeta | \btau & \sim \prod_{g=1}^G \prod_{j \in \G(g)} \mathcal{N} \left(0,\frac{1}{\lambda' _g \lambda_2} \frac{\tau_j - 1}{\tau_j} \right), \\
		\btau & \sim \prod_{j=1}^p \mathcal{TG} \left( \frac{1}{2},\frac{8 \lambda_2}{\lambda_1^2}, \left(1,\infty \right) \right).
		\end{align}
	\end{subequations}
	Here, $\mathcal{TG} \( k,\theta,\( x_l,x_u\) \)$ denotes the truncated gamma distribution with shape $k$, scale $\theta$, and domain $\(x_l,x_u\)$. In this Bayesian formulation the penalty parameters in $\blambda = \begin{bmatrix} \lambda_1 & \lambda_2 & \lambda'_1 & \cdots & \lambda'_G \end{bmatrix} \tr$ play the role of the hyperparameters in a Bayesian hierarchical model.
	
	\section{Estimation}\label{seq:estimation}
	\subsection{Empirical Bayes}\label{seq:empiricalbayes}
	If the penalty parameters are known, estimation of the frequentist elastic net model parameters or finding the posterior of the generalized elastic net model is feasible with small adjustments of the available algorithms \cite[]{friedman_regularization_2010,zou_regularization_2005} or MCMC samplers \cite[]{li_bayesian_2010}. Determining these penalty parameters, however, is not straightforward. 
	
	In the frequentist elastic net without group-wise penalisation, two main strategies are used: (i) estimate both $\lambda_1$ and $\lambda_2$ by cross-validation over a two-dimensional grid of values or (ii) re-parametrise the problem in terms of penalty parameters $\alpha= \frac{\lambda_1}{2 \lambda_2 + \lambda_1}$ and $\lambda = 2 \lambda_2 + \lambda_1$, fix the proportion of $L_1$-norm penalty $\alpha$ and cross-validate the global penalty parameter $\lambda$. Strategy (i) is advised by \cite{waldron_optimized_2011}, while (ii) is proposed by \cite{friedman_regularization_2010}. In the generalised elastic net setting, strategies (i) and (ii) imply $2 + G$ and $1 + G$ penalty parameters, respectively. $K$-fold cross validation over $D$ values then results in $K \cdot D^{2 + G}$ and $K \cdot D^{1 + G}$ models to estimate. Typically $K$ is set to 5, 10, or to the number of samples $n$, while $D$ is in the order of $100$, so that even for small $G$, the number of models to estimate is prohibitively large. 
	
	In the Bayesian framework, estimation of penalty parameters may be avoided by the addition of a hyperprior to the model hierarchy. The hyperprior takes the uncertainty in the penalty parameters into account by integrating over them. This approach introduces two issues. Firstly, the choice of hyperprior is not straightforward. Many authors suggest a hyperprior from the gamma family of distributions \cite[]{alhamzawi_bayesian_2017,mallick_bayesian_2013,kyung_penalized_2010}, but the precise parametrisation of this gamma prior is not so obvious. A second issue is the loss of correspondence between the  Bayesian and frequentist elastic net. If the ultimate goal is feature selection, this correspondence may be exploited through the automatic feature selection property of the frequentist elastic net. Endowing the penalty parameters with a hyperprior obstructs their point estimation and, consequently, impedes automatic feature selection. Therefore, to circumvent the problem of hyperprior choice and allow for feature selection by the frequentist elastic net, we propose to estimate the penalty parameters by empirical Bayes.
	
	Many forms of empirical Bayes exist, the most formal one being maximisation of the marginal likelihood with respect to the hyperparameters. The resulting hyperparameter estimates are then plugged into the prior. The marginal likelihood is often introduced as a measure of model evidence given the observed data and is computed by integrating the product of likelihood and prior with respect to the model parameters. In the case of the elastic net introduced in (\ref{eq:enmodel2}) this results in the following marginal empirical Bayes posterior for $\bbeta$:
	\begin{align}
	p_{\hat{\blambda}}(\bbeta | \y) & = \frac{\L (\y ; \bbeta) \pi_{\hat{\blambda}} (\bbeta) }{p_{\hat{\blambda}}(\y)} = \frac{\int_{\btau} \L (\y ; \bbeta) \pi_{\hat{\blambda}}(\bbeta | \btau) \pi_{\hat{\blambda}}(\btau) \, d\btau}{p_{\hat{\blambda}}(\y)}, \nonumber \\
	\hat{\blambda} & := \underset{\blambda}{\argmax} \log p_{\blambda} (\y) = \underset{\blambda}{\argmax} \int_{\bbeta} \int_{\btau} \L_{\blambda}(\y, \bbeta, \btau) \, d\bbeta d\btau \nonumber \\
	& = \underset{\blambda}{\argmax} \int_{\bbeta} \int_{\btau} \L (\y ; \bbeta) \pi_{\blambda}(\bbeta | \btau) \pi_{\blambda}(\btau) \, d\bbeta d\btau. \label{eq:mmlestimate}
	\end{align}
	The integrals in (\ref{eq:mmlestimate}) are intractable in the case of the elastic net. In the omics setting the integrals are also high-dimensional, in which case numerical and Monte Carlo approximation methods become tedious and computationally expensive. Moreover, Laplace approximation is known to suffer from low accuracy in many high-dimensional settings \cite[]{shun_laplace_1995}. In \cite{casella_empirical_2001} an EM algorithm is described that estimates the hyperparameters. This EM algorithm iteratively maximises the expected joint log likelihood, such that the sequence:
	\begin{equation}\label{eq:ebEM}
	\blambda^{(k + 1)} = \underset{\blambda}{\argmax} \E_{\bbeta, \btau | \y} \[ \log \L_{\blambda}(\y, \bbeta, \btau) | \blambda^{(k)} \]
	\end{equation}
	converges to a local maximum of the marginal likelihood. The difficulty herein is in the calculation of the expected joint log likelihood. \cite{casella_empirical_2001} suggests to approximate the expectation by its Monte Carlo expectation. Although elegant and simple, this method requires a converged MCMC sample from the posterior for every iteration: a computationally intensive procedure. We propose to tackle this problem by approximating the expectation in (\ref{eq:ebEM}) using variational Bayes.
	
	\subsection{Variational Bayes}\label{sec:variationalbayes}
	Variational Bayes is a widely used method to approximate Bayesian posteriors. It has successfully been applied in a wide range of applications, including genetic association studies \cite[]{carbonetto_scalable_2012} and gene network reconstruction \cite[]{leday_gene_2017}. In variational Bayes, the posterior is approximated by a tractable form and estimated by optimizing a lower bound on the marginal likelihood of this model (see Section 3 of the SM for the lower bound of the proposed model). For an extensive introduction and concise review, see \cite{beal_variational_2003} and \cite{blei_variational_2017}, respectively.
	
	To simplify the computations of our variational approximation, we follow \cite{polson_bayesian_2013} and introduce latent variables $\omega_i$, for $i=1, \dots, n$. Conditional on $\bbeta$, the $\omega_i$ are independent of the $y_i$ and P\'{o}lya-Gamma distributed (see Section 4 of the SM for more details). We augment Model (\ref{eq:enmodel2}) with:
	\begin{equation}\label{eq:logmodel}
	\bomega | \bbeta \sim \prod_{i=1}^n \mathcal{PG}\(m_i, |\x_i \tr \bbeta| \).
	\end{equation}
	Our variational Bayes approximation to the posterior distribution of (\ref{eq:enmodel2}) and (\ref{eq:logmodel}) factorizes over blocks of parameters. We choose the blocks such that:
	\begin{equation}\label{eq:varbayesapproximation}
	p (\bomega, \bbeta, \btau | \y) \approx Q(\bomega, \bbeta, \btau) = q_{\bomega} (\bomega) q_{\bbeta} (\bbeta) q_{\btau} (\btau).
	\end{equation}
	Writing $\btheta_1 = \bomega$, $\btheta_2 = \bbeta$, $\btheta_3 = \btau$, and $\btheta = \begin{bmatrix} \btheta_1 & \btheta_2 & \btheta_3 \end{bmatrix}$, calculus of variations gives the optimal distributions $q^*_{\btheta_j} (\btheta_j) \propto \exp \{\E_{\btheta \backslash \btheta_j} [\log p (\btheta | \y)]\}$, where optimality is achieved in terms of the Kullback-Leibler divergence of the posterior to the approximate distribution \cite[]{neal_view_1998}. The approximation in (\ref{eq:varbayesapproximation}) renders both the posterior parameter calculations and the expected joint log likelihood as introduced in (\ref{eq:ebEM}) tractable.
	
	After a change of variables $\psi_j = \tau_j - 1$, we find the optimal distributions in our variational Bayes implementation for the model parameters as:
	\begin{equation}\label{eq:varbayesdistr}
	q^*_{\bbeta} (\bbeta) \sim \mathcal{N} (\bmu, \bSigma) \text{, } q^*_{\bomega} (\bomega) \sim \prod_{i=1}^n \mathcal{PG} (m_i, c_i)\text{, and } q^*_{\bpsi}(\bpsi) \sim \prod_{j=1}^p \mathcal{GIG} \(\frac{1}{2}, \frac{\lambda_1^2}{4 \lambda_2}, \chi_j\), 
	\end{equation}
	where $\mathcal{GIG} (\cdot)$ denotes the generalized inverse Gaussian distribution (See SM Section 5 for the derivations). The so-called variational parameters in (\ref{eq:varbayesdistr}) contain cyclic dependencies, so we update them by:
	\begin{subequations}\label{eq:VBupdateequations}
		\begin{align}
		\bSigma^{(t + 1)} &= \( \X \tr \bOmega^{(t)} \X + \lambda_2 \bLambda' + \frac{\lambda_1 \sqrt{\lambda_2}}{2} \bLambda' \Z^{(t)}\)^{-1}, \\ & \text{with } \bOmega^{(t)} = \diag\[ \(\frac{m_i}{2 c_i^{(t)}}\) \tanh \( \frac{c_i^{(t)}}{2} \)\] \text{ and } \Z^{(t)}=\diag\[(\chi_j^{(t)})^{-1/2}\], \nonumber \\
		\bmu^{(t + 1)} &= \bSigma^{(t + 1)} \X \tr (\y - \m/2), \\ 
		c_i^{(t + 1)} &= \sqrt{\x_i \tr \bSigma^{(t + 1)} \x_i + (\x_i \tr \bmu^{(t + 1)})^2} \text{, for } i=1, \dots, n, \\
		\chi_j^{(t + 1)} &= \lambda'_{g(j)} \lambda_2 \[\bSigma^{(t + 1)}_{jj} + (\bmu^{(t + 1)}_j)^2 \] \text{, for } j=1, \dots, p,
		\end{align}
	\end{subequations}
	until convergence. Here, $\bLambda'$ is a diagonal matrix with entries $\lambda'_g$, each repeated $|\G(g)|$ times and $\m = \begin{bmatrix} m_1 & \cdots & m_n \end{bmatrix} \tr$. Furthermore, $\mathbf{A}_{jj}$ and $\mathbf{a}_j$ denote the $j$th diagonal element of a square matrix and $j$th element of a column vector, respectively. Naive calculation of the variational parameters is computationally expensive. In Section 6 of the SM we show that informed calculation of the parameters results in a significant reduction of computational complexity.
	
	\subsection{Empirical-variational Bayes}\label{seq:empvarbayes}
	Variational Bayes was shown to underestimate the posterior variance of the parameters, both numerically and theoretically, in several settings \cite[]{rue_approximate_2009,consonni_mean-field_2007,bishop_pattern_2006,wang_inadequacy_2005}. This coincides with our experience that the global penalty parameters $\lambda_1$ and $\lambda_2$ tend to be overestimated. To prevent overestimation we use the parametrisation of the elastic net in \cite{friedman_regularization_2010} as introduced in Section \ref{seq:empiricalbayes}: we fix $\alpha$ and estimate $\lambda$ by cross-validation of the regular elastic net model, such that the overall penalisation is determined by cross-validation of only $\lambda$. By combining cross-validation of the global penalty parameter $\lambda$ with empirical Bayes estimation of the penalty multipliers $\blambda' = \begin{bmatrix} \lambda'_1 & \cdots & \lambda'_G \end{bmatrix} \tr$, the estimation is more robust to underestimation of the variational posterior variances. The remaining issue is the choice of $\alpha$. \cite{hastie_glmnet_2016} recommend to either fix $\alpha$ \textit{a priori}, or compare the results for several choices of $\alpha$. We recommend the latter. Otherwise we note that, in our experience, the choice $\alpha=0.5$ often gives good results.
	
	For estimating the penalty multipliers, the intractable posterior expectation in (\ref{eq:ebEM}) is approximated by the variational posterior:
	$$
	\E_{Q} \[ \log \L_{\blambda'}(\y, \bomega, \bbeta, \btau) | \blambda'^{(k)} \] = \frac{1}{2} \sum_{g=1}^G |\G(g)| \log (\lambda'_g) - \frac{(1 - \alpha) \lambda}{4} \sum_{g=1}^G \lambda'_g d^{(k)}_g + C,
	$$
	where $C$ is constant in $\blambda'$ (see SM Section 7 for the full derivation). The $d_g^{(k)}$ terms are calculated as: $d^{(k)}_g = \sum_{j \in \mathcal{G} (g)} \[\bSigma^{(k)}_{jj} + (\bmu^{(k)}_j)^2\]\(1 + \alpha \lambda^{1.5} \sqrt{\frac{1 - \alpha}{8 \chi_j^{(k)}}}\)$. An estimate of the new penalty multipliers is now given by:
	\begin{subequations}\label{eq:mmlupdateequation}
		\begin{align}
		\blambda'^{(k+1)} & = \underset{\blambda'}{\argmax} \frac{1}{2} \sum_{g=1}^G |\G(g)| \log (\lambda'_g) - \frac{(1 - \alpha) \lambda}{4} \sum_{g=1}^G \lambda'_g d^{(k)}_g \\
		& \text{subject to } \prod_{g=1}^G (\lambda'_g)^{|\G(g)|} = 1.
		\end{align}
	\end{subequations}
	Although the solution to (\ref{eq:mmlupdateequation}) is not available in closed form, this convex problem is easily solved by a numerical optimisation routine. The full procedure is summarized in Section 8 of the SM.
	
	\section{Extensions}\label{sec:extensions}
	\subsection{Unpenalized covariates}
	Inclusion of an intercept $\beta_0$ is achieved by appending the data matrix $\X$ with a column of ones. Penalization of such an intercept parameter is not desirable \cite[]{Hastie_Elements_2009}. Additionally, it is often desirable to include unpenalized covariates in the model. Common examples of such  covariates in clinical research are patient characteristics, such as age, BMI, and sex. In the following we assume the intercept to be included in the unpenalized covariates.
	
	To include unpenalized covariates, we divide the model matrix into two parts $\X = \begin{bmatrix}\X_u & \X_r\end{bmatrix}$, where $\X_u$ are the $u$ unpenalized variables and $\X_r$ are the $r$ penalized variables. Let $\bLambda'_*$ and $\Z_*$ be the matrices $\bLambda'$ and $\Z$ prepended with $u$ zero columns and $u$ zero rows. Then we have for the current estimate of the covariance matrix $\bSigma$:
	\begin{align}\label{eq:unpencov}
	\bSigma &=\(\X \tr \bOmega \X + \lambda_2 \bLambda'_* + \frac{\lambda_1 \sqrt{\lambda_2}}{2} \bLambda'_* \Z_*\)^{-1} \nonumber \\
	& = 
	\begin{bmatrix} 
	\X_u \tr \bOmega \X_u & \X_u \tr \bOmega \X_r \\
	\X_r \tr \bOmega \X_u & \X_r \tr \bOmega \X_r + \lambda_2 \bLambda' + \frac{\lambda_1 \sqrt{\lambda_2}}{2} \bLambda' \Z
	\end{bmatrix}^{-1}.
	\end{align}
	With the choice of blocks as in (\ref{eq:unpencov}), blockwise inversion renders the largest required matrix inverse $(\X_r \tr \bOmega \X_r + \lambda_2 \bLambda' + \frac{\lambda_1 \sqrt{\lambda_2}}{2} \bLambda' \Z)^{-1}$, of dimension $p \times p$. Inversion of this matrix is done efficiently by applying the Woodbury identity.
	
	\subsection{Monotonicity of the penalty parameters}
	Enforcing monotonicity of the penalty multipliers is desirable in some settings. An example of such a setting is if we have $p$-values from a previous, related study available. A $p$-values-based partitioning of the variables may, \textit{a priori}, be expected to render penalty multipliers that increase monotonically with $p$-value. That is, larger $p$-values may be expected to yield at least as large penalty multipliers as variables with smaller $p$-values. We propose to include this \textit{a priori} assumption by requiring the penalty multipliers to increase monotonically with $p$-value, thereby also stabilizing their estimates.
	
	A natural way of enforcing monotonicity is to extend the constraint in (\ref{eq:mmlupdateequation}) with $\lambda'_1 \leq \dots \leq \lambda'_G$. The problem is still convex and may be numerically solved. From experience, however, we note that in combination with this constrained optimisation, the EM algorithm described in Section~\ref{seq:empiricalbayes} often converges to a local optimum close to the initialisation. We therefore enforce monotonicity through a \textit{post hoc} isotonic regression on the penalty multipliers after every optimisation step.
	
	\subsection{Feature selection}
	Feature selection is often desirable in high-dimensional prediction problems and omics data problems are no exception. A selection of biomarkers may lead to a large cost reduction by supporting targeted assays. Bayesian feature selection is often done by inspection of posterior credible intervals. Similarly as in frequentist hypothesis testing, we may select a feature if zero is not contained in the credible interval. However, the Bayesian lasso's credible intervals (a special case of the elastic net) are known to suffer from low frequentist coverage in sparse settings \cite[]{castillo_bayesian_2015}. We often assume sparsity of the features in omics research, so selection by credible intervals is generally inconsistent. We therefore propose to select features in the frequentist paradigm.
	
	As shortly touched upon in Section \ref{seq:empiricalbayes}, frequentist feature selection is trivial after estimation of the penalty multipliers. We simply plug the estimated penalty parameters into some frequentist elastic net algorithm that allows for differential penalization. In our own package \texttt{gren}, we use the \texttt{R}-package \texttt{glmnet} \cite[]{friedman_regularization_2010}. In the frequentist elastic net, feature selection is then done automatically. Furthermore, to select a specific number of features, we simply adjust the global $\lambda$ until we select the desired number of features.
	
	\subsection{Multiple partitions}
	In many cases the features may be partitioned in more than one way. For example, we may have both information on annotation and $p$-values from a previous study available. A naive way of incorporating multiple partitions is to cross-tabulate the partitions and create a separate group for every combination. This poses two problems: (i) The number of penalty parameters increases exponentially with the number of partitions and (ii) some of these combinations may contain only few features, so that the estimation procedure becomes unstable. We propose to stabilise the procedure and keep the number of parameters to estimate manageable by modelling the penalty parameters multiplicatively. 
	
	We describe our implementation here for two partitions of the features. To this end, let $(\G_1(1), \dots, \G_1(G_1))$ and $(\G_2(1), \dots, \G_2(G_2))$ denote the two partitions, containing $G_1$ and $G_2$ groups respectively. Furthermore, in the following we assume that the empty sum and empty product evaluate to 0 and 1, respectively. In this two-partition setting we have two penalty multipliers per feature, represented by $\lambda'_{g_1}$ and $\lambda''_{g_2}$, respectively. The generalised frequentist elastic net estimator and corresponding conditional prior are now:
	\begin{subequations}\label{eq:multpart}
		\begin{align}
		\hat{\bbeta} &:= \underset{\bbeta}{\argmax} \ell(\y ; \bbeta) -  \frac{\lambda_1}{2} \sum_{g_1=1}^{G_1} \sum_{g_2=1}^{G_2} \sqrt{\lambda'_{g_1} \lambda''_{g_2}} \sum_{\mathclap{\substack{j \in \G_1(g_1) \\ \cap \G_2(g_2)}}} | \beta_j| - \frac{\lambda_2}{2} \sum_{g_1=1}^{G_1} \sum_{g_2=1}^{G_2} \lambda'_{g_1} \lambda''_{g_2} \sum_{\mathclap{\substack{j \in \G_1(g_1) \\ \cap \G_2(g_2)}}} \beta_j^2, \\
		\bbeta | \btau & \sim \prod_{g_1=1}^{G_1} \prod_{g_2=1}^{G_2} \prod_{\substack{j \in \G_1(g_1) \\ \cap \G_2(g_2)}} \mathcal{N} \left(0,\frac{1}{ \lambda'_{g_1} \lambda''_{g_2} \lambda_2} \frac{\tau_j - 1}{\tau_j} \right).
		\end{align}
	\end{subequations}
	After switching to the parametrisation in \cite{friedman_regularization_2010}, the new penalty multiplier estimates $\blambda'= \begin{bmatrix} \lambda'_1 & \cdots & \lambda'_{G_1} \end{bmatrix} \tr, \blambda''= \begin{bmatrix} \lambda''_1 & \cdots & \lambda''_{G_2} \end{bmatrix} \tr$ are given by:
	\begin{subequations}\label{eq:estmultpart}
		\begin{alignat}{2}
		\blambda'^{(k+1)}, \blambda''^{(k+1)} &:= \underset{\blambda', \blambda''}{\argmax} && \Bigg\{ \frac{1}{2} \sum_{g_1=1}^{G_1} |\G_1(g_1)| \log (\lambda'_{g_1}) + \frac{1}{2} \sum_{g_2=1}^{G_2} |\G_2(g_2)| \log (\lambda''_{g_2}) \\ & && - \frac{(1 - \alpha) \lambda}{4} \sum_{g_1=1}^{G_1} \sum_{g_2=1}^{G_2} \lambda'_{g_1} \lambda''_{g_2} d^{(k)}_{g_1 g_2} \Bigg\} \nonumber \\       & \text{subject to } && \prod_{g_1=1}^{G_1} \prod_{g_2=1}^{G_2} (\lambda'_{g_1} \lambda''_{g_2})^{|\G_1(g_1) \cap \G_2(g_2)|} = 1,
		\end{alignat}
	\end{subequations}
	with the $d_{g_1 g_2}$ terms calculated as $\sum_{\substack{j \in \G_1(g_1) \\ \cap \G_2(g_2)}} \[\bSigma^{(k)}_{jj} + (\bmu^{(k)}_j)^2\]\(1 + \alpha \lambda^{1.5} \sqrt{\frac{1 - \alpha}{8 \chi_j^{(k)}}}\)$. The optimisation in (\ref{eq:estmultpart}) is again a convex problem that is easily solved by some numerical optimisation routine. This method naturally generalises to more than two partitions of the data.
	
	In (\ref{eq:multpart}) we model the two partition-specific penalty multipliers in a multiplicative way, mainly for convenience. First, there are computational reasons: The separation of the square root in the $L_1$-norm penalty term into two terms facilitates the numerical estimation. Additionally, multiplicative modelling of the partitions allows for more flexibility in the estimates. To see this consider the obvious alternative: additive partition-specific multipliers. Since the penalty multipliers are strictly positive, the constraint on the geometric mean makes one large additive penalty estimate difficult to compensate with a smaller estimate, thereby impairing flexibility of the model. Large multiplicative penalty multipliers are easier to compensate for by smaller penalty multipliers, rendering this the less restrictive option for modelling the partitions.
	
	\section{Simulations}\label{sec:simulations}
	\subsection{Setup}\label{sec:setup}
	To assess variable selection and predictive performance we conduct a simulation study in which we compare \texttt{gren} to the regular elastic net and ridge models, and \texttt{GRridge} \cite[]{van_de_wiel_better_2016}. \texttt{GRridge} is similar to \texttt{gren} in the sense that it estimates group-specific penalty multipliers. The two main differences with \texttt{gren} are (i) the absence of an $L_1$-norm penalty and (ii) the estimation procedure. For the regular elastic net and \texttt{gren}, we fix the proportion of $L_1$-norm penalty to three different values: $\alpha \in \{0.05, 0.5, 0.95 \}$. The first setting closely resembles the ridge setting, where $\alpha=0$, the third one is similar to the lasso with $\alpha=1$. For the methods that do automatic variable selection (the regular elastic net and \texttt{gren}), we estimate the models for a range of model sizes. Likewise for \texttt{GRridge}, where we use the \textit{post hoc} variable selection method described by \cite{novianti_better_2017}.
	
	The quality of feature selection is assessed by Cohen's kappa \cite[]{cohen_coefficient_1960}. This kappa is measures similarity between two sets of categorical values. One set contains the feature selection indicators: $s(j) := \mathbbm{1}\{\hat{\beta}_j \neq 0\}$, while the other set contains the true non-zero feature indicators: $t(j) := \mathbbm{1}\{\beta_j \neq 0\}$. We calculate Cohen's $\kappa$ as:
	\begin{align*}
	\kappa & = \frac{f_o - f_e}{1 - f_e}, \text{ where } f_o = p^{-1}\sum_{j=1}^p \mathbbm{1}\{s(j)=t(j)\} \\ & \text{ and } f_e = p^{-2} \left\{ \(\sum_{j=1}^p s(j)\) \cdot \(\sum_{j=1}^p t(j)\) + \[\sum_{j=1}^p (1 - s(j)) \] \cdot \[\sum_{j=1}^p (1 - t(j)) \]\right\}.
	\end{align*}
	Here, $f_o$ and $f_e$ denote the frequencies of correctly identified and expected correctly identified features, respectively. A positive or negative kappa indicate better and worse feature selection than expected by chance, respectively. In addition, we evaluate estimation accuracy by mean squared error: $\text{MSE} = p^{-1} \sum_{j=1}^p (\beta_j - \hat{\beta}_j)^2$. 
	
	Predictive performance is measured by area under the receiver operator curve (AUC) and Brier skill score (BSS). AUC gives the area under the curve of sensitivity versus $1-$ specificity and equals the probability that the classifier ranks a randomly chosen case higher than a randomly chosen non-case. The BSS is a normalised version of the Brier score: $\text{BSS} = 1 - \frac{\sum_{i=1}^n (y_i - \hat{y}_i)^2}{\sum_{i=1}^n (y_i - \bar{y})^2}$,
	where $\hat{y}_i=\expit(\x_i \tr \hat{\bbeta})$ is the prediction for observation $i$ and $\bar{y}=n^{-1} \sum_{i=1}^n y_i$. The BSS measures the fraction of Brier score improvement compared to a non-informative model.
	
	We created a training set of $n=100$ samples to estimate the models and a test set of $n_{\text{test}}=1000$ samples on which we compute the performance measures. The features $\x_i$, $i=1, \dots, n$, are sampled from a $p$-dimensional Gaussian distribution centred at zero, where we set $p=1000$. We introduce correlation between the features through a block diagonal covariance matrix for the $\x_i$. The blocks are of sizes $25 \times 25$, with off-diagonals set to $\rho=0.7$ and diagonals to $\sigma^2 = 1$.
	
	The model parameters $\beta_j$, $j=1, \dots, p$, are simulated in four groups of 250 with differing signal strengths and are created in two steps. First, we draw 250 parameters from the elastic net distribution, parametrised as in \cite{friedman_regularization_2010}, where $\alpha=0.5$ and $\lambda=100$ in all groups, and different penalty multipliers, $\lambda' \in \{ 0.14, 0.51, 1.95, 7.39 \}$. The second step consists of setting the (in absolute value) smallest 125 parameters in a group to zero. This results in a total of 500 zero and 500 non-zero parameters, evenly distributed over the four groups.
	
	Finally, we simulate the outcome data $y_i$, $i=1, \dots, n$, from a binary logistic model: $y_i \sim \mathcal{B}(1, \expit (\x_i \tr \bbeta))$. To mitigate the influence of random variation in the simulations, we repeated every simulation 50 times and report estimated penalty parameters and LOESS smoothed averages of the performance measures for a range of model sizes.
	
	\subsection{Results}\label{sec:results}
\begin{figure}[!h]
		\centering
		\includegraphics[width=0.8\linewidth]{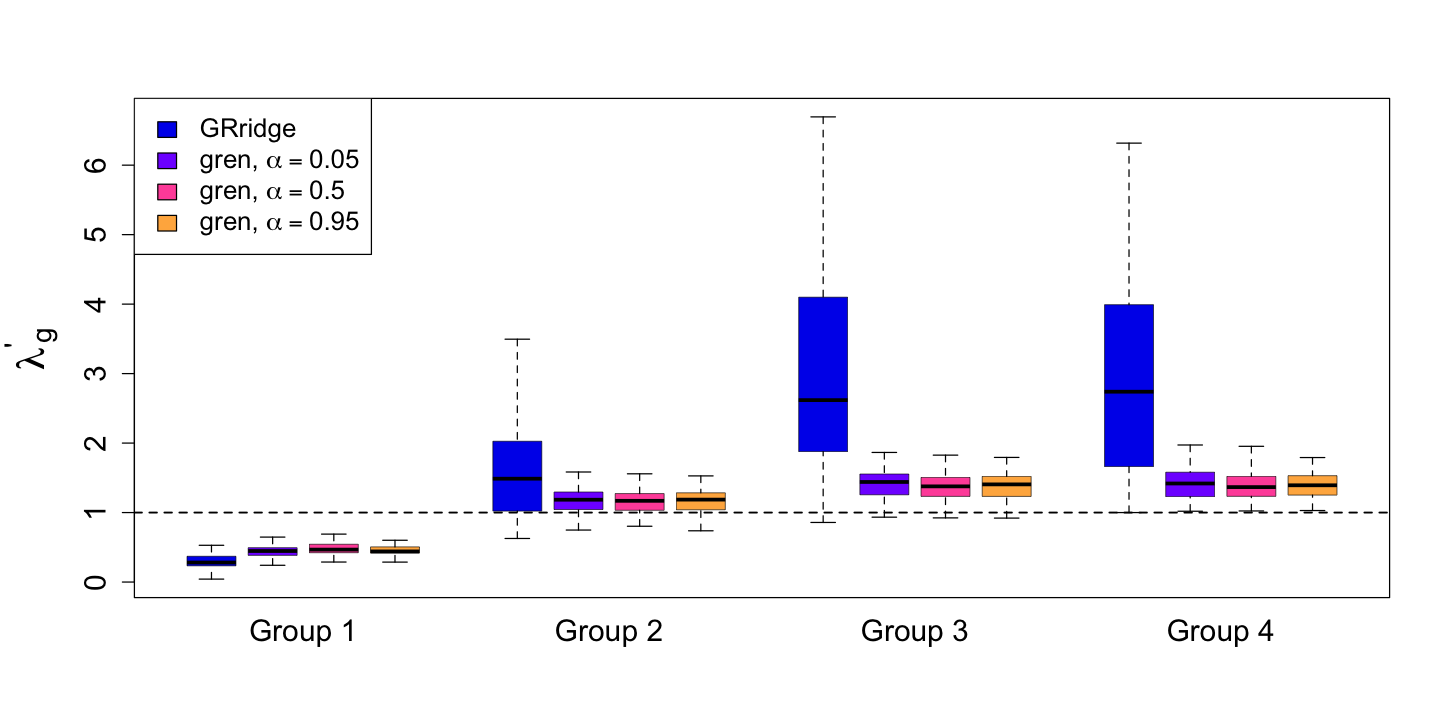}
		\caption{Estimated penalty multipliers of the group-regularized methods based on simulated data.}
		\label{fig:sim1_penalties}
	\end{figure}

	For all models, the estimated penalty parameters follow the expected pattern (Figure \ref{fig:sim1_penalties}); roughly, the penalty multipliers increase with decreasing signal strength. \texttt{GRridge}'s pattern of penalty multipliers is more pronounced than that of \texttt{gren}. However, both methods show difficulty in distinguishing between the two lower signal groups. In addition, we note that the \texttt{GRridge} penalty multipliers estimates show a lot more variation than \texttt{gren} in the three lower signal groups.
	
\begin{figure}[!h]
		\centering
		\includegraphics[width=1\linewidth]{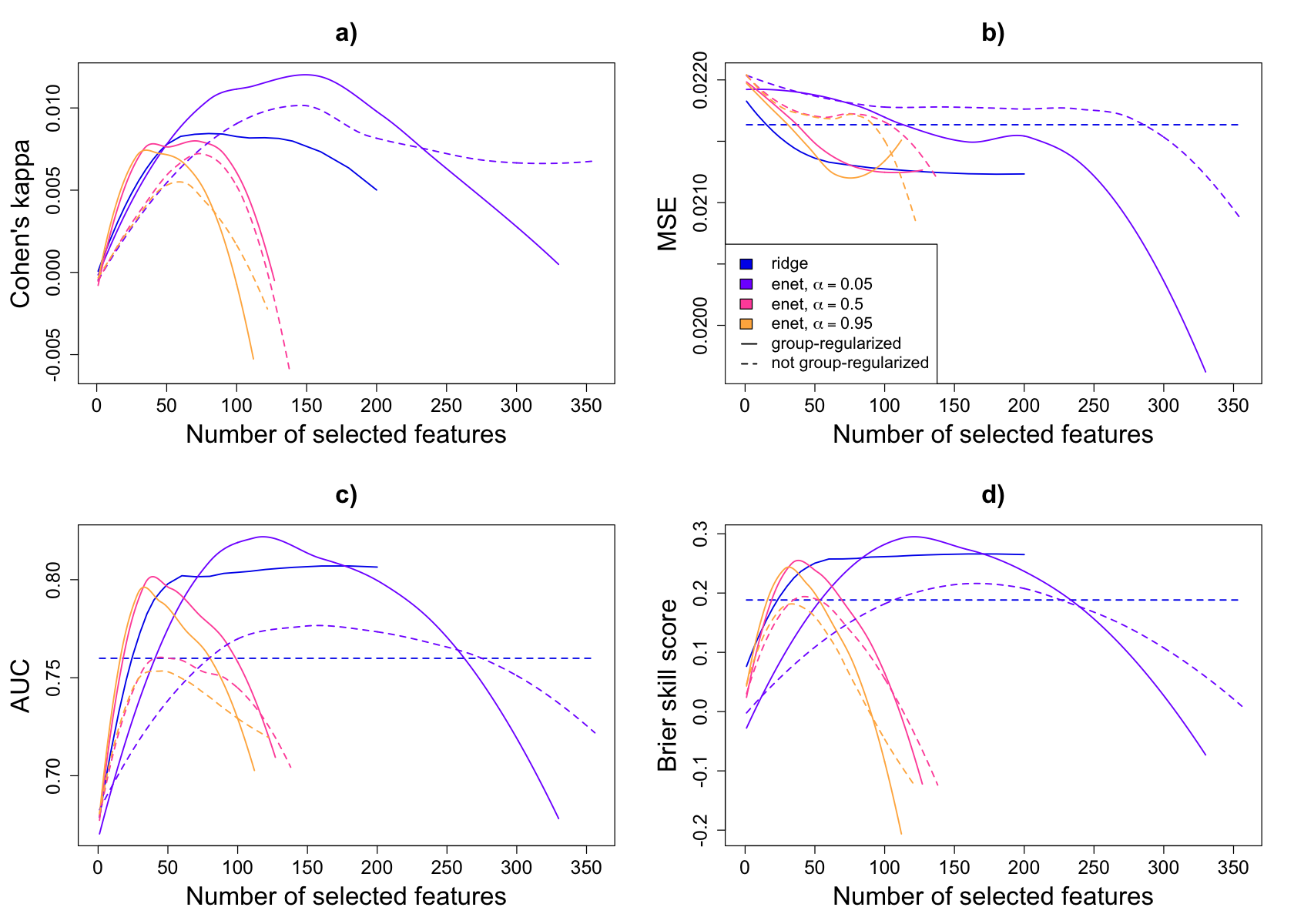}
		\caption{LOESS fit of (a) Cohen's kappa, (b) MSE, (c) AUC, and (d) BSS for (group-regularized) ridge and (group-regularized) elastic net with $\alpha \in \{ 0.05, 0.5, 0.95\}$ based on simulated data.}
		\label{fig:sim1_performace}
	\end{figure}

	For all performance measures, the group-regularized methods outperform the non group-regularized ones (Figure \ref{fig:sim1_performace}). Cohen's kappa shows that \texttt{gren} with $\alpha=0.05$ generally outperforms the other methods in terms of variable selection. \texttt{gren} with $\alpha=0.05$ is, however, less able to reconstruct the coefficients, evident from the generally higher MSE. In addition, SM Section 9 shows that \texttt{gren} strongly improves feature selection stability, compared to the regular elastic net.  
	
	For predictive performance, competitiveness of \texttt{gren} depends on the range of model sizes. For the smaller models, the \texttt{gren} models with $\alpha \in \{ 0.5, 0.95\}$ tend to outperform the other methods, while for the larger models, \texttt{GRridge} and \texttt{gren} with $\alpha=0.05$ tend to perform better. All elastic net methods seem to deteriorate in performance after a certain model size, because of the selection of `noisy' features after (most of) the predictive features have been selected. The more lasso-like models (with $\alpha \in \{0.5, 0.95\}$) prefer smaller models compared to \texttt{GRridge} and the elastic net with $\alpha=0.05$, a tendency already noted in the original elastic net paper \cite[]{zou_regularization_2005}. In all, when \texttt{gren} does not outperform the other methods, it's performance is on a par.
	
	\section{Application to microRNAs in colorectal cancer}\label{sec:app}
	\subsection{Partitioning based on differential expression}\label{sec:mirna}
	The example data set is from a deep sequencing analysis on microRNA \cite[]{neerincx_combination_2018}. The study was done in 88 treated colorectal cancer patients, with the aim of classifying treatment response, coded as either non-progressive/remission (70 patients) or progressive (18 patients). After pre-processing and normalisation, 2114 microRNAs remained. In addition to the 2114 microRNAs we incorporated 4 unpenalized clinical covariates into the analysis: prior use of adjuvant therapy (binary), the type of systemic treatment regimen (ternary), age, and primary tumor differentiation (binary).
	
	In a preliminary experiment on different subjects, the microRNA expression levels of metastatic colorectal tumour tissue were compared to normal non-colorectal tissue and primary colorectal tumour tissue was compared to primary colorectal tumour tissue \cite[]{neerincx_mir_2015}. This yielded 221 microRNAs that were differentially expressed in both comparisons ($\text{FDR} \leq 0.05$), versus 1893 not differentially expressed microRNAs. We expect that incorporation of this partitioning enhances therapy response classification, because tumor-specific miRNAs might be more relevant than non-specific ones. In addition, we divided the differentially expressed microRNAs even further into 127 highly differentially expressed microRNAs ($\text{FDR} \leq 0.001$) and 94 medium differentially expressed microRNAs ($0.001 < \text{FDR} \leq 0.05$).  We will refer to this second partitioning as the three-group setting, as opposed to the first, the two-group setting.
	
\begin{figure}[!h]
		\centering
		\includegraphics[width=0.83\linewidth]{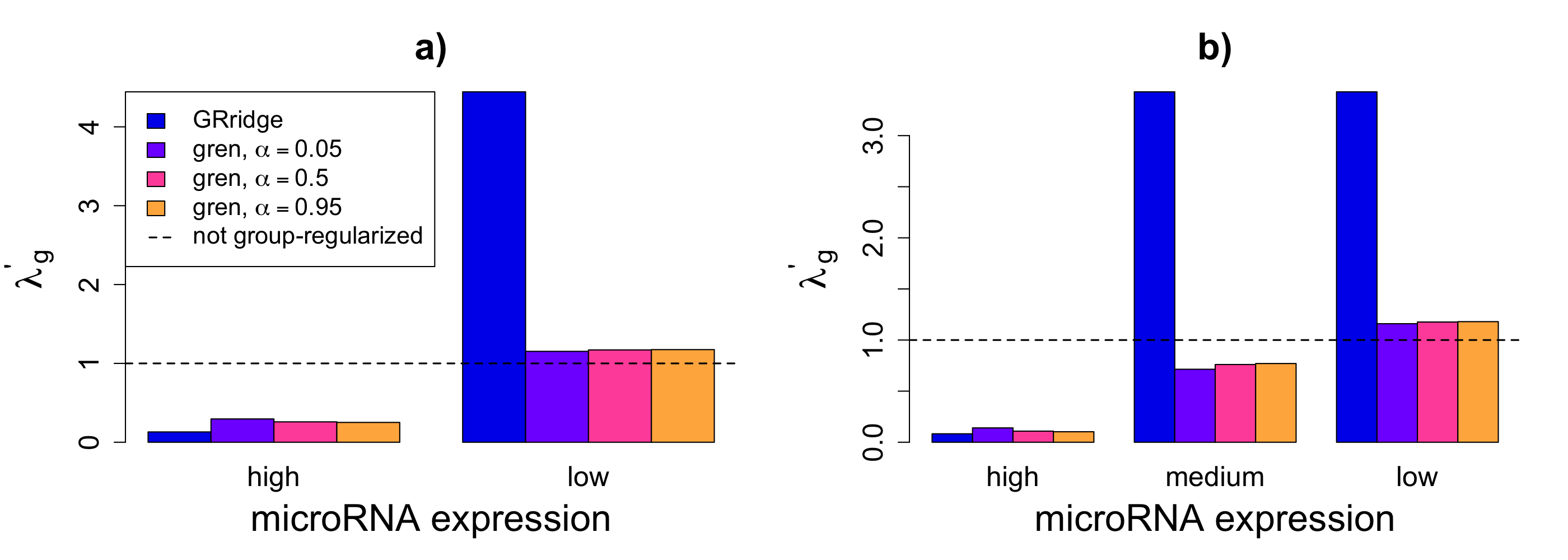}
		\caption{Estimated penalty multipliers for (a) the two-group setting and (b) the three-group setting.}
		\label{fig:col_bar}
	\end{figure}

	We compared \texttt{gren} to ridge and elastic net regression, and \texttt{GRridge}. For the elastic net methods (including \texttt{gren}) we set $\alpha \in \{ 0.05, 0.5, 0.95\}$. The estimated penalty multipliers are according to expectation: in all group-regularized methods, the 221 differentially expressed microRNAs receive the smallest penalty (Figure \ref{fig:col_bar}a). In the three-group setting the pattern is again as expected (Figure \ref{fig:col_bar}b): the highly expressed group, receives the lowest penalty, followed by the medium expressed group, while the non-expressed group receives the strongest penalty. \texttt{GRridge} is not able to distinguish between the medium and non-differentially expressed groups of microRNAs.
	
\begin{figure}[!h]
		\centering
		\includegraphics[width=1\linewidth]{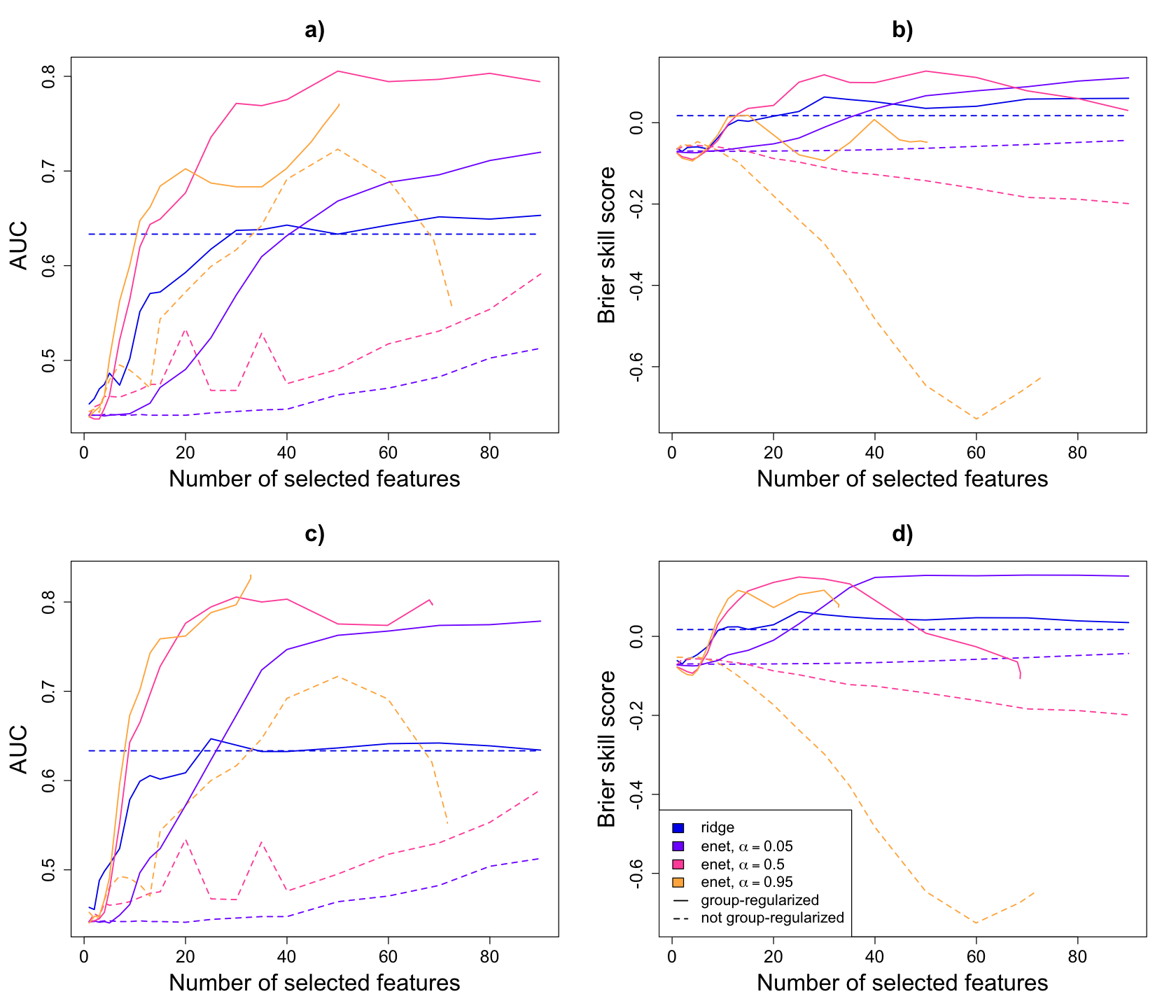}
		\caption{Cross-validated performance as (a) AUC in the two-group setting, (b) Brier skill score in the two-group setting, (c) AUC in the three-group setting, and (d) Brier skill score in the three-group setting against the number of selected features for each method.}
		\label{fig:col_per}
	\end{figure}

	Predictive performance of the methods is measured by AUC and BSS, both estimated by leave-one-out cross-validation (LOOCV), and shown in Figure \ref{fig:col_per}. All group-regularized elastic net models consistently outperform their non-group-regularized counterparts in both settings, in terms of AUC and BSS. In both settings, \texttt{gren} with $\alpha=0.5$ and $\alpha=0.95$ outperforms the other methods for a large range of model sizes in terms of AUC. This is especially true for the larger models. \texttt{gren} with $\alpha=0.5$ outperforms the other methods with respect to BSS in the smaller model ranges, while \texttt{gren} with $\alpha=0.05$ becomes competitive for the larger models. A general pattern in both settings is that \texttt{GRridge} and \texttt{gren} with $\alpha=0.05$ perform similarly. This is not surprising, because an elastic net model with $\alpha=0.05$ is close to a ridge model, in terms of penalisation. In general, the classifiers with three group penalties are better than the ones with two group penalties when relatively few features are used. Comparing, for example, the models of size 20 for \texttt{gren} with $\alpha=0.5$, we have an AUC of 0.68 for the two-group setting, versus an AUC of 0.78 for the three-group setting. Note that models with few features are often desirable for clinical implementation.
	
	\subsection{Random groups}
	Considering that the features may be partitioned into many groups, with one parameter per group, we have to be aware of overfitting risks. We investigated this using the data introduced in Section \ref{sec:mirna}, randomly dividing the features into three groups. We fixed the group sizes to the group sizes used in the three-group setting in Section \ref{sec:mirna}. Under this random partitioning of the features, we expect all penalty multipliers to be estimated as one if overfitting does not occur.
	
\begin{figure}[!h]
		\centering
		\includegraphics[width=0.5\linewidth]{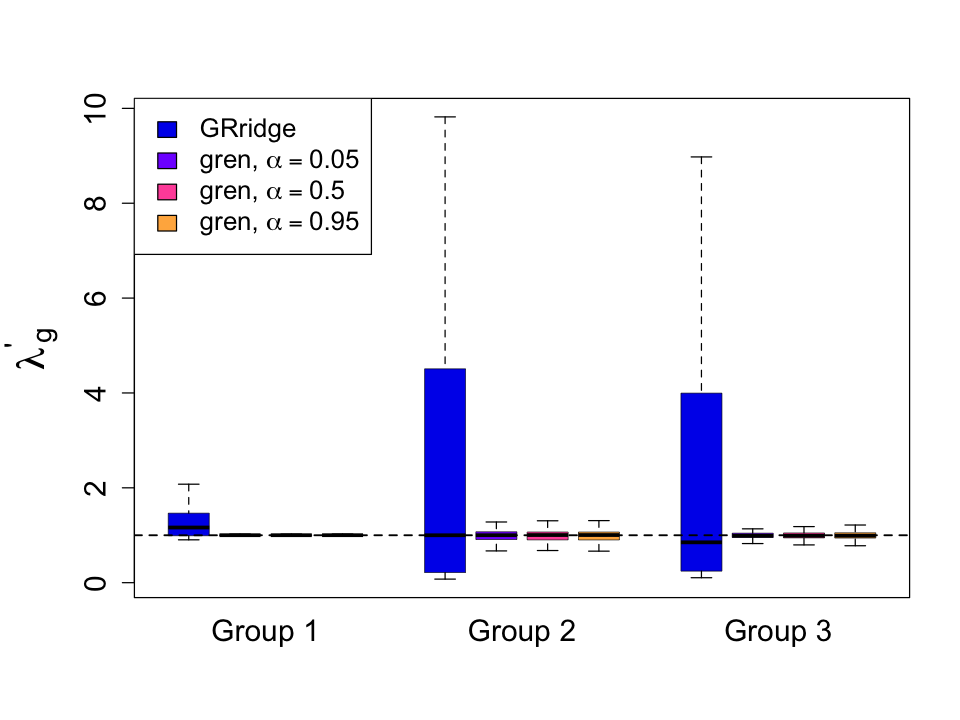}
		\caption{Estimated penalty multipliers of the group-regularized methods for 100 repeats of 3 random groups in colorectal miRNA data.}
		\label{fig:random}
	\end{figure}

	We compared the estimated multipliers to the estimates by \texttt{GRridge}. Since these results depend on one specific randomisation of the groups, we repeated the procedure 100 times and present the results in Figure \ref{fig:random}. From this figure we see that the estimates for \texttt{gren} are close to one. The estimates by \texttt{GRridge} show much more variation. Additionally, \texttt{GRridge} gives slightly biased estimates: the median penalty parameter estimates are 1.16, 1, and 0.85 for the three groups. In contrast, the median estimates of \texttt{gren} are 1, 1, and 0.99. We repeated the simulation with ten evenly sized groups and present the results in SM Section 9.
	
	\section{Discussion}\label{sec:discussion}
	In a taxonomy of Bayesian methods, the proposed method may be considered a local shrinkage model, as opposed to the global-local shrinkage priors that \cite{polson_local_2012,bernardo_shrink_2011} discuss. They characterise certain desirable properties of these global-local shrinkage priors in high dimensions, which, for example, the horseshoe possesses \cite[]{carvalho_horseshoe_2010,carvalho_handling_2009}. In our case, global shrinkage would imply adding another hyperprior for the global $\lambda_1$ and $\lambda_2$ (or $\alpha$ and $\lambda$) hyperparameters. We argue however, that if the groups are informative, the empirical Bayes estimation of the (semi-)global shrinkage parameters $\lambda'_g$ may be more beneficial than full Bayes shrinkage of the global penalty parameters, because the latter does not use any known structure to model the variability in the hyperparameters. Nonetheless, an interesting direction of future research is the extension of the group-regularized elastic net to a group-regularized horseshoe model, since the horseshoe has been shown to handle sparsity well and render better coverage of credibility intervals than lasso-type priors \cite[]{van_der_pas_horseshoe_2014}.
	
	Although our method can be considered weakly adaptive, it is different from the adaptive lasso \cite[]{van_de_geer_adaptive_2011,huang_adaptive_2008,zhang_adaptive_2007,zou_adaptive_2006} and adaptive elastic net \cite[]{zou_adaptive_2009} in the sense that it adapts to external information rather than to the data. It also differs in the scale of adaptation: in the adaptive lasso and elastic net the adaptive weights are feature specific, while in our case they are estimated on the group level, rendering the adaptation more robust. As can be seen from the simulations and real data example in Sections \ref{sec:simulations} and \ref{sec:app}, adaptation to external data may be beneficial for prediction and feature selection. We believe that this is due to the `borrowing of information' effect: estimates that are believed to be similar are shrunken in the same way, yielding overall, better estimates.
	
	Another obvious comparison is to the group lasso \cite[]{meier_group_2008,yuan_model_2006}. Although it is similar in the sense that it shrinks on the group level, the group lasso is built upon an entirely different philosophy: it sets whole groups of features to zero. The intended application of such a group-wise penalty is to small interpretable groups of features, like, for example, dummies of a categorical variable. Another difference between \texttt{gren} and the group lasso is the number of penalty parameters. \texttt{gren} estimates one parameter per group, while the group lasso estimates one overall penalty parameter; it is thereby less flexible in differential shrinkage of the parameters. In addition, in the simulations and data applications of \cite{van_de_wiel_better_2016}, group lasso prediction performed inferior to \texttt{GRridge} prediction.
	
	A possible weak point of the proposed method is the double EM loop. The double loop increases the chance of ending up in a local optimum. In the application discussed above we investigated the occurrence of multiple optima, but never encountered them. This does not guarantee that local optima do not occur, but it provides some evidence that local optima are not ubiquitous. Local optima can often be avoided by an informed choice of starting values. In our experience, reasonable starting values are obtained by running a group-regularized ridge regression and using the estimated penalty multipliers as starting values for \texttt{gren}. In addition, we note that even a local optimum might give `good enough' predictions in many practical settings.
	
	\section*{Software}
	The method is available as an \texttt{R} package from \url{https://github.com/magnusmunch/gren/}.
	
	\section*{Supplementary Material}
	Supplementary Material is available online from \url{https://arxiv.org}.
	
	\section*{Acknowledgements}
	\textit{Conflict of Interest}: None declared.
	
	\section*{Funding}
	This research has received funding from the European Research Council under ERC Grant Agreement 320637.
	
	\bibliographystyle{author_short3.bst}
	\bibliography{refs}
	
\end{document}